# Modelling polymer compression in flow: semi-dilute solution behaviour

*Dave E. Dunstan*

Department of Chemical and Biomolecular Engineering, University of Melbourne, VIC 3010, Australia.

 davided@unimelb.edu.au

ABSTRACT

Rheo-optical measurements on Polyemethyl methacrylate (PMMA) and 4-Butoxycarbonylmethylurethane (4-BCMU) show that these synthetic random coil polymer chains compress in Couette flow at semi-dilute concentrations. All previous models of polymer flow behaviour have assumed that the chains extend. A new model of polymers in flow in semi-dilute solution is presented where the chains compress in accord with the experimental findings. The chains are modeled as simple blobs exposed to a linear shear gradient which results in a compressive hydrodynamic force on the chain. Equating the hydrodynamic and elastic forces predicts chain compression with shear rate. The model also predicts a power law behaviour for polymer shear thinning of -2/3. The predicted power law for the decrease in radius with shear rate is -1/9 in close agreement with the measured value of -0.09+/-0.02 for both polymer systems.

INTRODUCTION

Polymers melts and solutions contribute a significant class of materials in both the synthetic and biological sciences.[1] Predicting the flow behaviour of polymeric solutions from the fundamental physics has long been the quest of soft condensed matter. An essential assumption of polymer dynamics is that the single molecule response to applied stress may be used to interpret the observed macroscopic material behaviour.[2-4] The elegant models of single polymer chains which assume that the chain can be described as a random walk on a periodic lattice have been successful in predicting a number of key properties. Furthermore, the inclusion of excluded volume to the ideal chain has enabled prediction of the solution size of polymers.[3] The entropy of the chain is derived in terms of the end-to-end vector. This model forms the basis of rubber elasticity and is used to incorporate elasticity in models of flow where chain deformation results in entropy reduction and elasticity. Furthermore, the *theory of rubber elasticity* that is based on similar physical assumptions predicts the elastic behaviour of rubbers over a wide strain range.[5] Importantly, the *"Rubber Theory"* predicts both the compressive and extensional behaviour of rubbers.[5] This agreement between the theory and experiment, albeit at effectively *"infinite"* molecular weight and high polymer concentration with excluded volume interactions neglected, gives confidence that the fundamental tenets of the theory are correct. However, due to their complexity there exist very few simulations of concentrated polymer solutions and melts.[6-9]

An assumption of all previous models of polymer flow is that the chains extend in response to the hydrodynamic forces.[6-9] Recent experimental evidence shows that synthetic polymer chains compress in Couette flow at semidilute concentrations.[10-12] Recent studies on semidilute DNA solutions show tumbling occurs.[13, 14] It appears that the general assumption of chain extension in flow may not be valid for concentrations above critical overlap in Couette flow.[10-13, 15] Furthermore, recent Brownian dynamics simulations for dilute solutions predict chain compression by neglecting excluded volume effects and including hydrodynamic interactions.[16, 17] While these simulations have been done for dilute chains, the neglect of excluded volume effects is consistent with cioncentrated solution behaviour. The inclusion of hydrodynamic interactions in concentrated solution where they are screened is not however consistent with the physics of concentrated solutions. Furthermore, it is worth noting that Kuhn originally proposed that both extensive and compressive hydrodynamic forces act on the chains as they tumble in Jeffrey orbits.[18, 19] In light of recent experimental evidence showing chain compression in Couette flow, a new approach is used here where the chains compress in response to the hydrodynamic forces. We also note that coil compression is an elastic event which leads to reduced friction in the system and is therefore consistent with observed shear thinning for polymer solutions.

RESULTS

The shear rate dependence of the end-to-end distance, r, has been measured for PMMA using FRET tagged chains in laminar Couette flow.[10] The conformation of BCMU in flow was measured using absorption spectroscopy where the change in segment length with shear is used to determine the change in polymer size.[12] The results are plotted on a log-log scale in Figure 1 below. Both polymers show a decrease in the end-to-end distance with increasing shear rate.

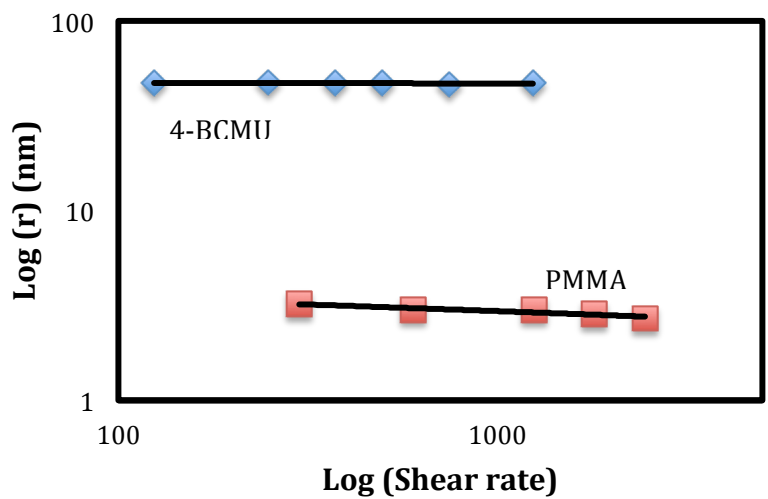

Figure 1. Measured end-to-end distance r versus shear rate. a. Data for 800kD 4-BCMU.[12] The fitted equation is: log r = -0.09log($\dot{\gamma}$) + 48.2 with the coefficient of determination: $R^2$ = 0.80. b. Data for 49kD FRET tagged PMMA in Couette flow.[10] The fitted equation is: log r = -0.09log($\dot{\gamma}$) + 4.4 with $R^2$ = 0.913. The lines of best fit yield an inverse 0.09+/-0.02 power of the radius with shear rate. The error bars are approximately the size of the symbols. The error associated with each point is: ~5% in the shear rate due to the radius/gap ratio of the Couette cell. For the 4-BCMU the unsheared size of the chain is 49+/-1nm and for PMMA the size is 4+/-0.1nm.

The results presented in Figure 1 are for two different rheo-optical experiments for two different polymeric systems. The data for PMMA was collected using time resolved FRET measurements on end tagged PMMA as a molecular tracer in a matrix of untagged PMMA at ~2C*.[10] The data for 4-BCMU is taken from reference 12 where the segment lengths of the 4-BCMU are measured to decrease with increasing shear rate at a polymer concentration of ~1.6C*.[12] Calculation of the average segment length and conversion to an end-to end distance using the equation $r = (Na^2)^{1/2}$ yields the results presented in Figure 1a. Here N is the number of segments and a the segment length as taken from literature values.[12] Interestingly, both data sets show a decreasing radius with shear rate with a power law of -0.09+/-0.02.

THEORY

In order to model the above data the polymer molecules are modeled as space filling blobs in the semi-dilute concentration range. The blobs behave as individual random chains that compress into themselves as the concentration is increased above critical overlap, (C*).[20] It has been recently shown that in the semi-dilute concentration range, mixing and comensurate entanglement is thermodynamically unfavourable.[20] The lowest free energy state of the chains is compressed into themselves and not interpenetrated and mixed as has been previously assumed. The thermodynamic interpretation suggests that entanglements are improbable. The non-entangled chains are then assumed to simply tumble and compress in response to the applied shear stresses as has been observed experimentally.[10-12] The blob is considered as approximated by a spherical object. Solc and Stockmeyer and independently Bruns were the first to recognise that the random chains are prolate objects.[21, 22] This has been recently shown in experimental rheo-optic measurements.[23] The spherical assumption is however adequate as the prolate objects of relatively small aspect ratio will be randomly oriented prior to flow.

For the semi-dilute solution the blobs experience both rotational and compressive forces in flow. The force acting on each half space in the Couette flow acts in the opposite direction and is simply one half of the Stokes' drag on the sphere. (See Figure 2) The hydrodynamic force on the sphere is assumed to be compressive. The resultant drag forces form a couple which tends to rotate the blob causing a tumbling motion and random orientation of the end-to-end vectors of the chains. This renders the compressive force on the chains to be isotropic relative to the end-to-end vectors of the ensemble of chains. We then assume that the the local hydrodynamic and elastic forces act along the velocity direction as shown in figure 2. The local forces may then be equated under steady state flow.

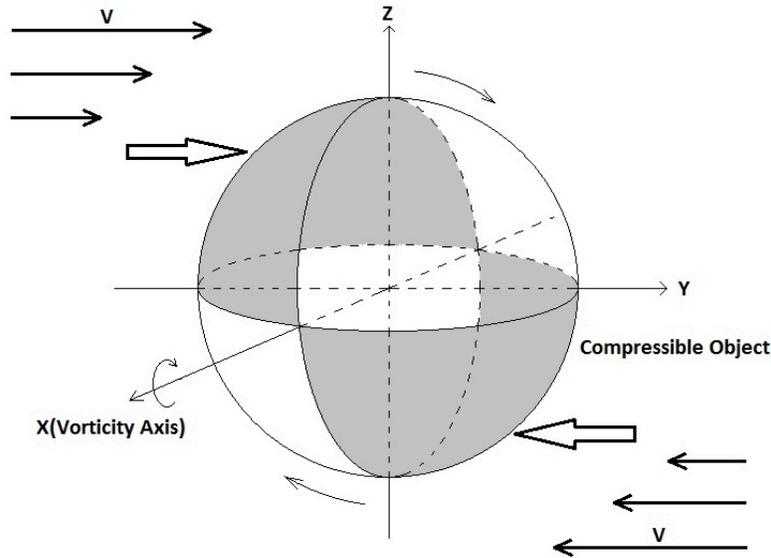

Figure 2. Schematic showing the polymer blob in Couette flow. The shaded area shows the region which experiences a compressive force from the flow. The upper half experiences a compressive force from left to right while the bottom right hand part of the image experiences a similar compressive force from the flow from right to left. Each surface experiences a compressive force equal to one half the Stokes' drag on a sphere. The total compressive force is then equal to $f_{compressive} = f_{hyd} = 6\pi\eta r v$ where η is the solution viscosity, r the sphere radius and v the velocity difference across the sphere in the direction shown. The sphere also experiences a torque around the vorticity axis. This rotational motion causes a tumbling which yields an averaged symmetric compression on the blobs in flow. At each point across the surface the hydrodynamic force is equal to the elastic force under steady shear. Local fluctuations will occur with the system reaching an average reduced size of the coil with increasing shear rate and commensurate hydrodynamic force. The arrows pointing inward on the blob represent the local hydrodynamic compressive forces.

Excluded volume effects are considered isotropic throughout solution and not incorporated in the model as previously discussed.[4] The macromolecules are approximated as ideal chains in concentrated solution as first proposed by Flory.[4]

The elastic and hydrodynamic forces on the ideal chain then act to change the dimensions of the coil in flow. The forces are used in the following treatment as at each point in the system the hydrodynamic and elastic forces oppose each other. In order for the system to reach steady state, the completely isotropic forces throughout the solution are equivalent. Obviously, the forces will fluctuate around the steady state average as the blobs rotate around the vorticity axis in the shear field. In steady state flow the hydrodynamic and elastic forces are then equated:

$$F_{el} = F_{hyd} \quad (1)$$

Where the magnitude of the elastic force for the ideal chain is a function of the end-to-end distance, r, as derived previously[24, 25]:

$$f_{el} = k_B T \left[ \frac{3r}{Na^2} - \frac{2}{r} \right] \quad (2)$$

Here $k_B$ is Boltzmann's constant, T the absolute temperature, r, the magnitude of the average end-to-end distance, N the number of freely jointed segments and a the segment length.[26] This form of the magnitude of the force law for the chains allows compression to occur. By defining the "Hookean" force law as previously used (first term in the brackets of equation 2), the chains are mathematically denied the ability to compress.[27] The second term in the brackets in equation 2 arises from treating the normalised distribution function for the probability of finding the ends of the chain in a spherical shell of radius r+dr. This form of the distribution function is the origin of the second term in equation 2. The form of the free energy equation then has a 2ln(r) term which gives rise to the 2/r term in the force equation. The magnitude of the elastic force embodies the correct physics in that at zero force the chain has finite size, which is the Flory size of the chain. The generally used form of the elastic force which includes only the first term in brackets in equation 2, the so called "Hookean" force law, requires that the chain size is zero at zero force. Neumann first posed the form given in equation 2 for the elastic force and recognised that it allowed compression to occur.[27] The form of the elastic force magnitude in Equation 2 correctly accounts for the finite size of the chain in the absence of external forces. The unit normal vector along the velocity direction multiplies the scalar magnitude of the elastic force in order that the vector forces are coupled. The elastic force then opposes the hydrodynamic force (grey arrow in Figure 2) along the velocity direction. We assume that the elastic and hydrodynamic forces act along the velocity direction. Under steady state flow, the hydrodynamic forces and the elastic forces may then be equated. (See Figure 2)

Multiplying Equation 2 by the unit normal vector to the surface of the blob returns the force to a vector which opposes the hydrodynamic force:

$$F_{el} = f_{el} . \tilde{n} \qquad (3)$$

At each localised point, we assume that the elastic force acting normal to the surface of the blob as shown in figure 2 opposes the hydrodynamic force.

The hydrodynamic force on the chain is:

$$F_{hyd} = 6\pi\eta\dot{\gamma}r^2 \qquad (4)$$

Here the viscosity of the polymer solution is $\eta$, $\dot{\gamma}$ the shear rate experienced by the chain and **r** the average end-to-end vector of the chain as defined above. It is assumed that the end-to-end distance is equivalent to the radius of the blob, which experiences the hydrodynamic force. The $r^2$ term arises from conversion of the velocity to the shear rate as $\dot{\gamma}$ = v/r so that the hydrodynamic force varies as $r^2$ in accord with the original derivation of the hydrodynamic drag on a dumbbell derived by Kuhn.[18] Kuhn's original paper shows that the chains may compress or extend. However, since Kuhn's original paper, extension has been assumed in the literature.[2, 3, 6, 7, 9, 28, 29] The overall hydrodynamic force on the blob has been experimentally shown to be compressive as shown in Figure 1. We now have that the elastic and hydrodynamic forces may be equated.

Combining [2] and [4] and using r = $(Na^2)^{1/2}$ (the usual Flory result for ideal chains to allow elimination of N and a) yields:

$$r^3 = \frac{k_B T}{6\pi\eta\dot{\gamma}} \tag{5}$$

Equation 4 indicates that the magnitude of the radius becomes unbounded at zero shear rate. This is not physically the case and suggests that the range of radius values must be restricted to those where the shear thinning is observed. This is a result of the substitution for $r = (Na^2)^{1/2}$ which is the zero shear value, to allow for a single power of the radius to be obtained. This substitution requires that there be a finite shear rate (change in r) for the values of r to be bounded. Over the shear-thinning region, equation 5 requires that the average end-to-end distance of the chains decreases with increasing shear rate as measured for the 4-BCMU and PMMA. The viscosity of the solution is a function of the polymer radius. The viscosity $\eta$ is not a constant in the system and will also reduce as the chains contract in size. It should be noted that the explicit solution for [5] only exists for chain compression, as an extensive hydrodynamic force is non-convergent.

To first order the viscosity of the solution, $\eta$, is approximated by a modified version of Einstein's equation:

$$\eta \sim \eta_0 \phi \tag{6}$$

Where the $\eta_0$ is the effective solvent viscosity and $\phi$ the volume fraction of the chains. We assume that the effective solvent viscosity composed of solvent and the surrounding polymers. As such it will depend on the polymer volume fraction and the shear rate. We assume that it is also proportional to the volume fraction $\phi$ so that:

$$\eta \sim \phi^2 \tag{7}$$

De Gennes and later Rubinstein and Colby have derived the same volume fraction dependence for the viscosity of semi-dilute solutions using scaling arguments. Furthermore, experimental data confirms the scaling arguments for polyethylene oxide in the semi-dilute concentration range.[3, 30] So that by assuming $\phi \sim r^3$ and by substitution into equation 5 we obtain the following:

$$\eta \sim [\dot{\gamma}]^{-\frac{2}{3}} \quad (8)$$

The generally accepted power law model is of the form:

$$\eta \sim [\dot{\gamma}]^{n} \quad (9)$$

Values reported for polymeric systems range between n~-0.2 to -1.[31-33] Interpretation of the data presented by Stratton indicates that for monodisperse polystyrene, n = -0.82.[33] The value of n = -2/3 predicted by the model is well within the range of accepted values for shear thinning polymers.[31]

The radius dependence with shear rate is then:

$$r \sim [\dot{\gamma}]^{-\frac{1}{9}} \quad (10)$$

DISCUSSION

The measured dependence of the decrease in the radius with increasing shear rate (power law of -0.09+/-0.02) is in close agreement with the model prediction of -1/9 (-0.11) (Equation 10). The close agreement for both polymers is remarkable and is strong evidence that the model posed embodies the key elements of the physics of Couette flow resulting in hydrodynamic forces on the polymer chains. Interestingly, the data for the two polymeric systems is determined from two different spectroscopic methods. The 4-BCMU curve is determined from measurements of the decreasing segment length in the chains while the data for the PMMA is determined from time resolved FRET measurements of the end-to-end distance of the chains in flow.

Furthermore, the model predicts a power law for the shear thinning viscosity of -2/3 (-0.67) that is within the range observed for polymer solutions which vary from -1.0 to -0.2.[31-33] Given that the model uses the approximation that the viscosity follows a volume fraction squared behaviour, the model appears an excellent fit to the power law behaviour of the radius and viscosity with shear rate. A correction to the viscosity-volume fraction dependence would presumably require higher order terms that would yield lower values of the predicted power law at higher concentrations. Indeed, scaling arguments predict that the viscosity goes as a 14/3 power of the volume fraction at concetrations above the entanglement concentration.[30] The measured viscosity-molecular weight behaviour for a range of polymers is consistent with the volume fraction dependence used in Equation 7.[34] Furthermore, de Gennes and later Rubinstein and Colby have modeled the viscosity-polymer volume fraction dependence described in equation 7 using scaling arguments.[3, 30] This relationship between the volume fraction of the polymer and the viscosity enables the macroscopic viscosity to be predicted as a function of the chear rate as in Equation 8. The predicted and measured decrease in radius with shear rate are in excellent agreement given the use of the second order dependence of the viscosity on volume fraction. It should be noted that the equations used are only valid in the range of shear thinning where the radius and viscosity are decreasing functions of the shear rate. Equation 5 yields an unbounded radius (and viscosity) as the shear rate tends to zero so that a modified form of the above equations must be used at low shear rates. This results from the substitution for $r = (Na^2)^{1/2}$ which then requires that the equations are only valid for the shear thinning range. The form of the equations at low shear rates will be similar to the Cross equation for shear thinning compared to the power law model.[32] Furthermore, the unresolved issue of shear induced phase separation observed in semi-dilute polymer solutions is explained by chain compression in flow. The observed compression in flow lays the foundation for an explanation of the observed shear induced phase changes observed for polymer solutions.[35, 36]

Furthermore, it is noted that the model predicts a value of $n = -1/2$ and a radius dependence of the shear rate with a power of -1/6 for dilute solutions where it is assumed that the viscosity is proportional to the volume fraction. To the best of our knowledge this has not yet been measured experimentally.

CONCLUSIONS

The model for polymers in flow is presented where the chains behave as elastically deformable blobs that compress in simple shear flow at semi-dilute concentrations. Simply equating the elastic and hydrodynamic forces on the blob enables the power law behaviour observed for shear thinning and the reduction in end-to-end distance with shear rate to be predicted over the shear thinning range. Physically the model is consistent with the observed rheology of polymer solutions which is attributed here to compression of the chains in flow rather than the previously assumed extension. Development of the model using the assumption that the chains compress will enable simple prediction of polymer visco-elastic behaviour including the power law for shear thinning.

ACKNOWLEDGEMENTS

I would like to thank the referees for their insightful comments that significantly improved the manuscript. I would also like to thank Lee White for stimulating discussions.